\documentclass[iop,apj,12pt,numberedappendix]{emulateapj}
\usepackage{times}
\usepackage[normalem]{ulem}
\usepackage{graphicx}
\usepackage{natbib}
\usepackage{amsfonts}
\usepackage{amsmath}
\usepackage{multirow}
\usepackage{enumerate}
\usepackage{comment}
\usepackage{float}
\usepackage{verbatim}
\usepackage[para]{threeparttable}
\usepackage[usenames,dvipsnames]{xcolor}
\usepackage[plainpages=false, colorlinks=true, anchorcolor=blue, linkcolor=blue, citecolor=blue, bookmarks=false]{hyperref}
\usepackage{scrextend}
\newcommand{\be}{\begin{eqnarray}}
\newcommand{\ee}{\end{eqnarray}}

\newcommand{\shortauth}{Morozova et al.}
\newcommand{\slugcom}{Submitted for publication in The Astrophysical Journal Letters}
\slugcomment{\slugcom}

\newcommand{\GG}[1]{}

\lefthead{\sc \footnotesize \slugcom \hfill \shortauth}
\righthead{\sc \footnotesize \slugcom \hfill \shortauth}

\begin{document}

\title{The influence of late-stage nuclear burning on red supergiant
supernova light curves}

\author{Viktoriya Morozova\altaffilmark{1}}
\author{Anthony L. Piro\altaffilmark{2}}
\author{Jim Fuller\altaffilmark{3}}
\author{Schuyler D. Van Dyk\altaffilmark{4}}
\altaffiltext{1}{Department of Physics, The Pennsylvania State University, University Park, PA 16802-6300, USA;
  vzg5138@psu.edu}
\altaffiltext{2}{The Observatories of the Carnegie Institution for Science, 813 Santa Barbara St., Pasadena, CA 91101, USA}
\altaffiltext{3}{TAPIR, Walter Burke Institute for Theoretical Physics, Mailcode 350-17, Caltech, Pasadena, CA 91125, USA}
\altaffiltext{4}{Caltech/Spitzer Science Center, Caltech/IPAC, Mailcode 100-22, Pasadena, CA 91125, USA 0000-0001-9038-9950}

\begin{abstract}

Many Type II supernovae (SNe) show hot early ($\sim$30 days) emission, and a 
diversity in their light curves extending from the Type IIP to the Type IIL, 
which can be explained by interaction with dense and confined circumstellar material 
(CSM). We perform hydrodynamical simulations of red supergiants to model the 
ejection of CSM caused by wave heating during late-stage nuclear burning. 
Even a small amount of deposited energy ($10^{46}-10^{47}$ erg), which is 
roughly that expected due to waves excited by convection in the core, is sufficient 
to change the shapes of SN light curves and bring them into better agreement 
with observations. 
As a test case, we consider the specific example of SN 2017eaw, which shows 
that a nuclear burning episode is able to explain the light curve if it occurs 
$\sim$150-450 days prior to core-collapse. Due to the long timescale it takes for 
the low energy shock to traverse the star, this would manifest as a pre-SN 
outburst $\sim$50-350 days prior to the full-fledged SN.
Applying work like this to other SNe will provide a direct connection between 
the SN and pre-SN outburst properties, which can be tested by 
future wide field surveys. In addition, we show that our models can qualitatively
explain the short lived `flash-ionization' lines seen in the early spectra
of many Type II SNe.

\end{abstract}

	
  
\section{Introduction}
\label{sec:introduction}

The pre-explosion images of hydrogen-rich, plateau type supernovae
(SNe~IIP) unambiguously identify red supergiants (RSGs) as their progenitors
\citep{smartt:09,vandyk:17}. Yet numerical models of SN~IIP 
light curves performed with up-to-date RSG models from stellar evolution codes still
cannot reproduce some of the basic observed features, such as fast rise and
early maxima of the light curves and the bright emission over the first
$\sim$$10-30$ days. 
\citep{morozova:17,morozova:18,moriya:17,moriya:18,paxton:18}. These studies show that addition
of a dense circumstellar material (CSM) on top of the RSG models can help
to bring them in a better agreement with the observations and may explain the full
diversity from Type IIP (plateau) to IIL (linear) subclasses. From the physical
point of view, this may be attributed to the fact that the CSM introduces a 
distinctly shorter diffusion timescale into the model (a few tens of days, 
compared to a $\sim$100$\,{\rm d}$ timescale of the bulk of the RSG envelope),
while the larger radius leads to less adiabatic cooling and brighter emission.
However, the underlying origin of the dense CSM remains a mystery. 
Models in which the CSM is produced by a 
dense wind require mass loss rates of at least $10^{-3}\,M_{\odot}\,{\rm yr}^{-1}$,
and it is not clear if such high rates are physically plausible.

A possible answer to this question is the ejection of matter by
a RSG at the late stages of its evolution. The energy required for the
ejection can be produced by vigorous late-stage core nuclear burning
processes, and deposited into the envelope by means of wave transport
\citep{quataert:12,shiode:14,quataert:16,fuller:17}. This scenario implies that
the luminosity of the progenitor star should change in the years prior to the
SN explosion, either in a steady manner or in the form of an outburst.

To date, there has been no unambiguous detection of an outburst
preceding a regular SN~IIP or IIL (here, we do not discuss Type IIn SNe, for which
pre-explosion outbursts have been detected \citealt{ofek:13,ofek:14}). 
Studies of the available pre-explosion images \citep{kochanek:17,johnson:17,oneill:18}
report no significant variability of the SN~IIP progenitors in the last several years
of their lives. At the same time, these images are often sparsely sampled and rarely
cover the infrared bands, which contain the largest fraction of the 
RSG emission. Interpretation of the pre-explosion
data is challenged by the fact that we do not yet have a
clear idea about the strength, time and duration of the pre-explosion outbursts that 
would be sufficient to explain the observed SN~IIP light curves.

In this Letter, we attempt to clarify these questions by drawing a 
more consistent picture of the influence of pre-explosion
wave energy transport on SN light curves.
Using numerical simulations, we model the hydrodynamics of the
matter ejection
caused by wave energy transport in a parameterized way. 
We vary the amount of energy deposition in the stellar envelope, and
the time between envelope heating and the core-collapse explosion. This way, we 
generate a progression of light curves that would correspond to
consequent snapshots of the progenitor profile after the outburst, showing that
the pre-explosion ejection of matter changes the SN light curves in a way 
that improves their agreement with the observations. Using an example
of a well-observed typical IIP SN 2017eaw, we deduce the
most likely values of the outburst energy and time between the
outburst and the core-collapse, finding they are similar to expectations
of wave heating models.

In Section~\ref{setup}, we describe our numerical setup. Section~\ref{results}
contains our main results, while Section~\ref{conclusions} is devoted to the
conclusions and discussion.


\section{Numerical Setup}
\label{setup}

We perform our study using the publicly available code 
\texttt{SNEC} \citep{morozova:15}. 
The code solves Lagrangian hydrodynamics coupled with
radiation transport in the flux-limited diffusion approximation.
This approximation is well suited for obtaining the bolometric light
curves of SNe~II starting from 1-2 days after the shock breakout and 
until the end of the plateau phase. We work with a solar metallicity,
$15\,M_{\odot}$ (at zero-age main sequence, ZAMS) stellar evolution
model from the KEPLER set by \citet{sukhbold:16}, evolved to the
pre-collapse RSG stage. This model is in 
agreement with the bolometric luminosity 
$(1.2\pm0.2)\times 10^{5}\,L_{\odot}$ deduced from the pre-explosion images of 
SN 2017eaw by \citet{vandyk:19}\footnote{\citet{vandyk:19} used Geneva evolutionary
tracks \citep{georgy:13} to estimate the ZAMS mass of the progenitor as
$\approx15\,M_{\odot}$, but we arrive to the
same conclusion based on the pre-explosion luminosities of KEPLER models.}.
The final mass of the RSG at the onset of core collapse is $12.6\,M_{\odot}$, while its radius
is $841\,R_{\odot}$.

In the first stage of our simulations, we model the deposition of energy by 
convectively excited waves during a vigorous late-stage nuclear burning episode 
into the RSG envelope (it could be core Ne or O burning), as outlined 
by \citet{fuller:17}. However, instead of modeling the wave
transport between the core and the envelope self-consistently, 
we parameterize our setup in terms of the energy injected at the base of
the hydrogen envelope. To deposit the energy into the model, we first 
excise a large part of its core, down to the 
density values of $\sim1\,{\rm g}\,{\rm cm}^{-3}$ 
($4.31\,M_{\odot}$ in terms of the excised inner mass). 
This enables us to 
simulate a very weak energy deposition with a reasonably large numerical time
step, while still keeping the inner boundary well inside the stellar core. 
In the current version of SNEC, the velocity at the inner boundary is always
taken to be zero.
We employ a commonly used thermal bomb mechanism, but instead of 
putting the energy into the innermost grid points, we inject it all in a 
single grid point at the density of $\sim 7\times 10^{-6}\,{\rm g}\,{\rm cm}^{-3}$. 
The duration of the thermal bomb is chosen so that the heating rate is equal to
$10^7\,L_{\odot}$, as in Figure 4 of \citet{fuller:17}. We vary the injected energy,
$E_{\rm inj}$, within the range of values expected in a standard RSG  
(see Figure 5 of \citealt{fuller:17}), namely, between $1\times 10^{46}\,{\rm erg}$
and $2\times 10^{47}\,{\rm erg}$, in steps of $0.5\times 10^{46}\,{\rm erg}$. 

The deposition of energy initiates a weak shock wave propagating through the
envelope of the RSG. Once this shock wave reaches the surface of the star, some
part of the envelope material gets ejected and the model starts expanding. 
Depending on the value of $E_{\rm inj}$, not all of the material ejected by the weak
shock wave becomes unbound, and part of it may later fall back onto the star.
We follow the evolution of the model for $\approx$900 days and collect the 
density, temperature and velocity profiles at different times after the energy 
injection, $t_{\rm inj}$. 

In the second stage of our simulations, we explode these 
profiles in SNEC, this time using our regular core-collapse SN setup
\citep{morozova:15,morozova:18}. As a result, we obtain
a two-dimensional grid of SN light curves corresponding to different energies $E_{\rm inj}$
and times $t_{\rm inj}$ between the energy injection and the core-collapse. 

Since this study uses as an example a well-observed SN 2017eaw, we 
restrict ourselves to a single value of $^{56}$Ni mass, $M_{\rm Ni}=0.075\,M_{\odot}$,
which was found to fit the radioactive tail of this SN in \citet{vandyk:19}
\footnote{\citet{buta:19} obtained a higher $^{56}$Ni mass of $0.115\,M_{\odot}$
for the same SN.}. 
In our models, we mix radioactive $^{56}$Ni up to $5\,M_{\odot}$ in mass coordinate.
In addition, we use a single value of the explosion energy (parameter 
$E_{\rm fin}$ in SNEC), which fits the plateau part of SN 2017eaw,
$0.75\times 10^{51}\,{\rm erg}$. The thermal
bomb energy in SNEC is computed as $E_{\rm bomb}=E_{\rm fin}-E_{\rm init}$,
where $E_{\rm init}$ is the initial energy of the model. 
Before exploding the models, we reattach their C/O cores back, assuming that
they have not been affected by the weak energy injection. Then, we excise
the inner $1.6\,M_{\odot}$ of material that forms a neutron star, and
deposit the thermal bomb energy into the first few grid points with the total mass of
$\Delta M = 0.02\,M_{\odot}$ for a duration of $1\,{\rm s}$
\footnote{After reattaching the cores, we do not attempt to smooth the
density profiles at the core-envelope interface. This leads to artificial bumps
at the transition between plateau and radioactive tails in the final SN light 
curves (see Figure~\ref{fig:fig4}). In reality, Rayleigh-Taylor instabilities
occurring when the shock wave propagates through the sharp density gradients
will smooth the profiles and, consequently, the light curves (see, for example, 
\citealt{paxton:18}). However, this effect is not important for the conclusions of
the present study, and we leave it for future work.}. We compare the resulting
light curves to the data of SN 2017eaw and find the best fitting model by
minimizing $\chi^2$.

\begin{figure}
  \centering
  \includegraphics[width=0.45\textwidth]{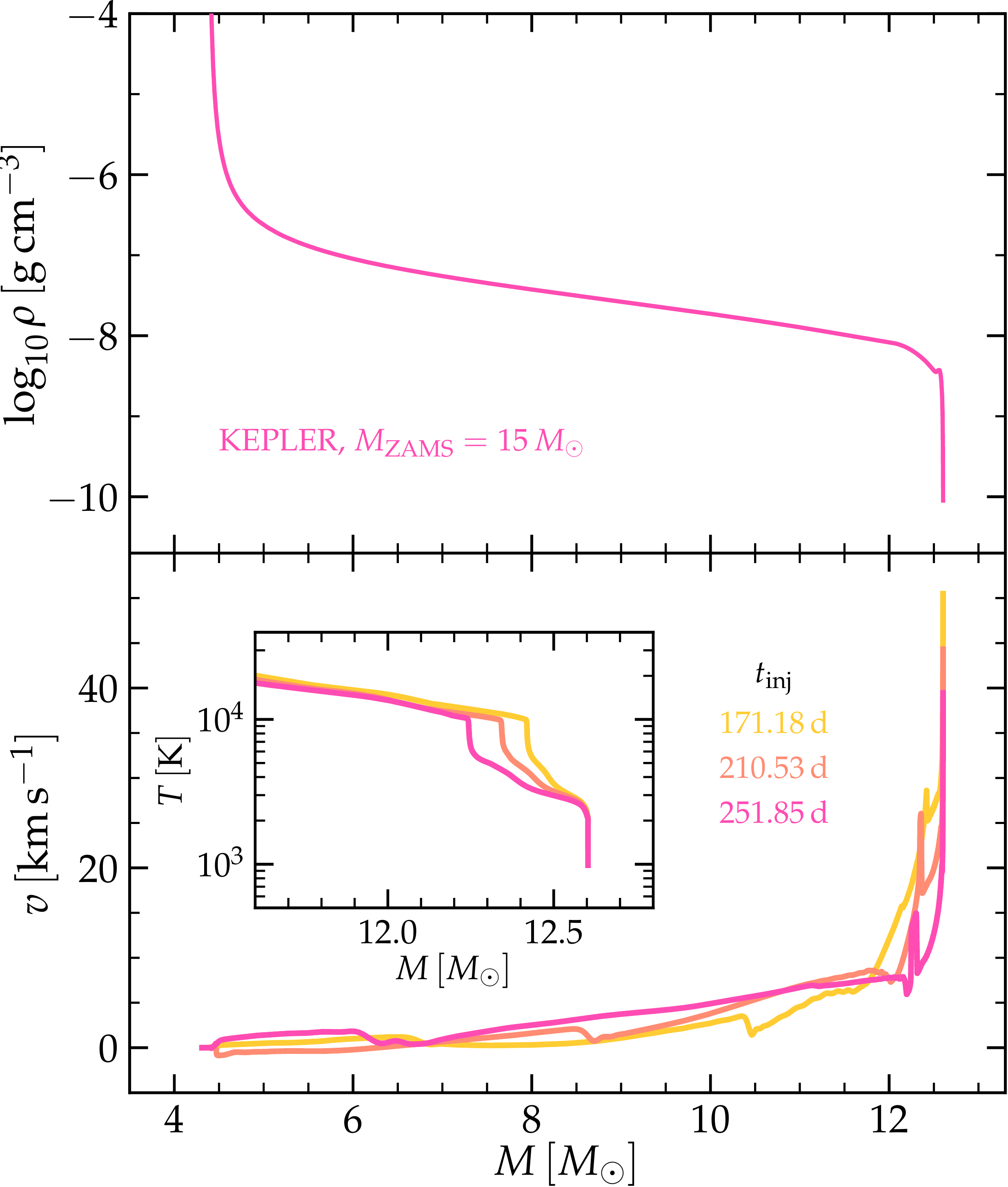}
  \caption{Top panel: RSG density profile used in our study.
  We mimic the energy input from the late-stage core nuclear burning 
  by injecting the chosen amount of energy, $E_{\rm inj}$, at the base of the 
  hydrogen envelope (in the model shown here, 
  we inject $5\times 10^{46}\,{\rm erg}$ 
  of energy at the point where the density is 
  $\approx 7\times 10^{-6}\,{\rm g}\,{\rm cm}^{-3}$). Bottom 
  panel: Velocity profiles at the different times after the energy injection. 
  The inset shows the corresponding temperature profiles in the outer regions of the 
  envelope.} 
  \label{fig:fig1}
\end{figure}

The main limitation of our study comes from the fact that SNEC does not have
a nuclear burning routine and a prescription for the convective energy transport, 
which are essential for supporting the stellar
structure in the evolutionary codes. It does not cause any problems in
quick and energetic SN explosions, which SNEC was originally designed to
simulate. However, following a stellar evolution model without these
physical components for several hundreds of days between the pre-SN
outburst and the core-collapse causes distortions in the outer layers 
of the model. This leads to artificial
bumps and irregularities in the pre-explosion velocity and density profiles 
(like in the bottom panel of Figure~\ref{fig:fig1}, and in 
Figure~\ref{fig:fig2}), which are otherwise expected to be smoother
\footnote{In fact, these bumps may in part be real and attributed to the wave
heating.~However in our current setup it is not possible to separate 
the real physical effect from the numerical artifacts.}.
Nevertheless, we argue that SNEC is able to capture the hydrodynamics
of the outburst, the near-surface density profile, and to give a robust qualitative prediction
of its influence on the final SN light curves. The aim of this work
is to draw the connection between the late-stage nuclear burning and
the shape of the SN~IIP light curves, as well as to get a general idea of the 
energetics and timing of the pre-explosion outbursts. In the future, we
plan to perform a more extended study of the outbursts with an improved
setup.


\section{Results}
\label{results}

\subsection{Effect on pre-SN structure}

\begin{figure}
  \centering
  \includegraphics[width=0.41\textwidth]{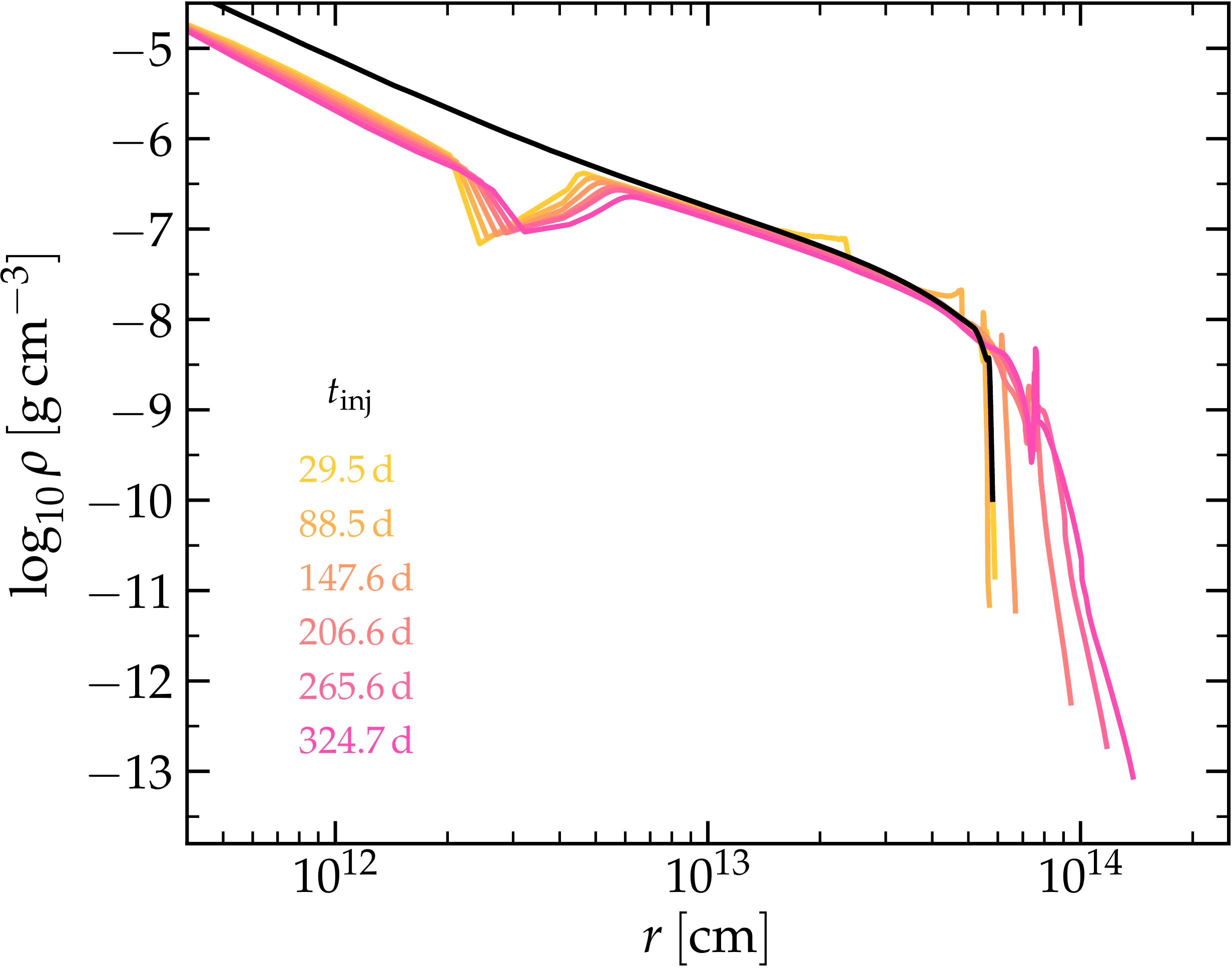}
  \caption{Evolution of the density profile of the RSG during the first few 
  hundreds of days
  after the energy has been injected at the base of its hydrogen envelope.
  The injected energy is $E_{\rm inj}=5\times 10^{46}\,{\rm erg}$.
  Black curve shows the model before the energy injection.
  The low density regions at the radial coordinate $\sim 3\times10^{12}\,{\rm cm}$ roughly
  correspond to the energy injection regions, while the irregularities at 
  $\sim8\times10^{13}\,{\rm cm}$ are due to the distortion of the stellar evolution progenitors 
  in SNEC (see last paragraph of Section~\ref{setup}).} 
  \label{fig:fig2}
\end{figure}

Figures~\ref{fig:fig1} and~\ref{fig:fig2} illustrate the evolution of the 
RSG profile in the first few hundreds of days after the energy has
been injected at the base of its hydrogen envelope 
($E_{\rm inj}=5.0\times 10^{46}\,{\rm erg}$ in both figures). 
The top panel of 
Figure~\ref{fig:fig1} shows the original pre-explosion density profile of the 
KEPLER model. 
The bottom panel of Figure~\ref{fig:fig1}
shows the velocity and temperature profiles of this model at 
different moments of time after the energy injection. Note that the time
it takes for a weak shock wave to propagate through the hydrogen
envelope all the way to the surface is $\sim$100 days 
in this model, a significant fraction of the time between energy
injection and core-collapse.

The velocity evolution in the bottom panel of Figure~\ref{fig:fig1}
reflects the pre-explosion density structure of the progenitor RSG.
The RSG model has a
shallower density profile in the bulk of its envelope, which transitions
into a steeper profile at about $0.5\,M_{\odot}$ below the surface
(top panel of Figure~\ref{fig:fig1}).
As a consequence, the velocity gained by the outermost $\sim 0.5\,M_{\odot}$
of the envelope after the passage of the weak shock wave 
launched by energy injection is noticeably 
higher than the velocity of the inner envelope regions. After the initial expansion, part of 
this outer material decelerates and at later times starts falling back 
towards the star.

The inset in the bottom panel of Figure~\ref{fig:fig1}
illustrates the temperature evolution in this model. The expansion of the
ejected material quickly cools it down to temperatures below 
$\sim 6000\,{\rm K}$. At this temperature,
the hydrogen and helium recombine, which leads to a dramatic drop in the
opacity of the stellar material. 
Therefore, by day $\sim 250$ the outermost $0.4-0.5\,M_{\odot}$ of the
model from Figure~\ref{fig:fig1} are effectively 
transparent to the optical emission. 
This is consistent with Figure 11 of \citet{fuller:17}, where the pre-heated
models have up to a few tenths of a solar mass of material
above the photosphere after the outburst.
At the same time, this material does not have enough time to cool down to
temperatures below $\sim 1000\,{\rm K}$, at which it would start to
form dust and obscure the progenitor. In the model shown 
in Figure~\ref{fig:fig1},
only the last few grid points reach the temperatures below $10^3\,{\rm K}$.
These cells contain the amount of mass of the order of $10^{-4}\,M_{\odot}$, and they are
likely affected by the boundary conditions used in SNEC. 
Similarly, the material expelled due to wave heating in MESA models by
\citet{fuller:17} does not cool down to temperatures below $2000\,{\rm K}$
prior to core-collapse. For this reason, we
do not expect a significant dust formation in the first few hundreds of 
days after nuclear burning episode causing the wave heating, unless
the ejected material can cool down further through the mechanisms that are not taken 
into account in both codes (e.g., cooling by lines).

\begin{figure*}
  \centering
  \includegraphics[width=0.95\textwidth]{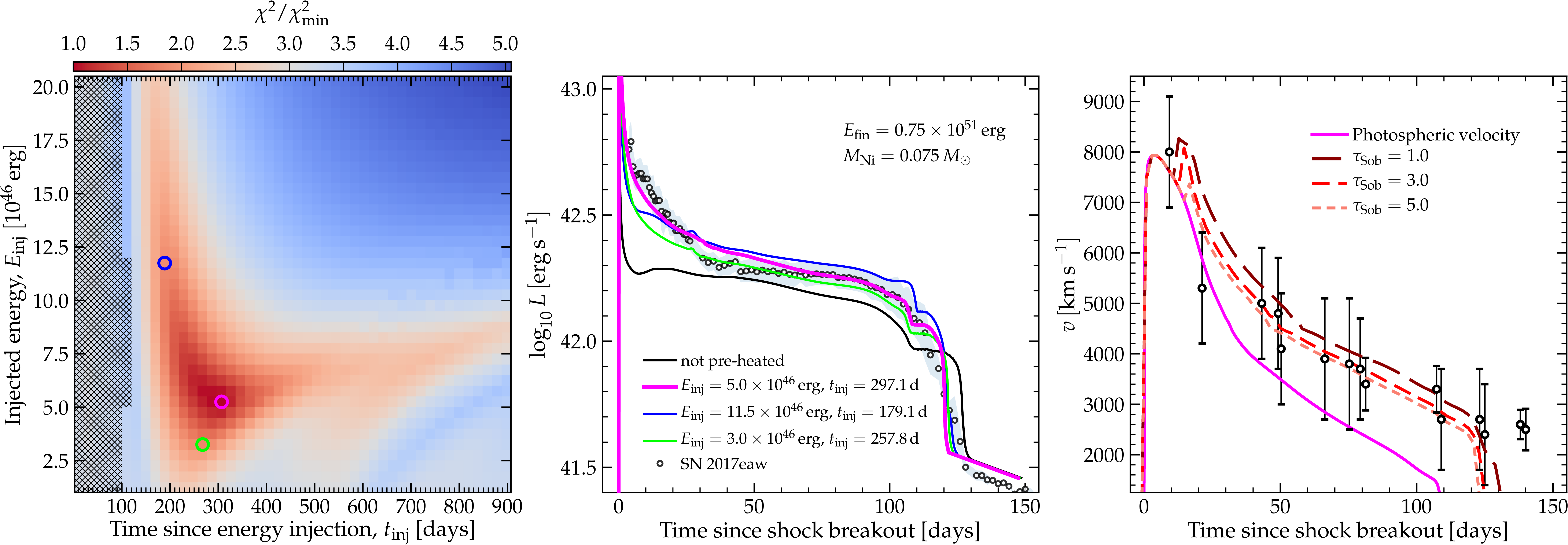}
  \caption{Left panel: Color coded distribution of $\chi^2/\chi^2_{\rm min}$ 
  across the $E_{\rm inj}$-$t_{\rm inj}$ parameter space for the KEPLER
  $15\,M_{\odot}$ model. The gray shaded region covers the times during which the weak
  shock wave launched by energy injection propagates between 
  the base of the hydrogen envelope and the stellar surface (in those times, 
  no visible outburst is expected from the progenitor). 
  The $\chi^2$ itself is determined by the comparison of the SN
  light curves obtained from these models
  to the bolometric light curve of SN 2017eaw. All SN explosions are performed
  with the same set of parameters,  
  $E_{\rm fin}=0.75\times 10^{51}\,{\rm erg}$ and $M_{\rm Ni}=0.075\,M_{\odot}$. The
  magenta circle corresponds to the minimum $\chi^2$, while the lime and blue
  circles provide examples of reasonably well fitting models with smaller and larger $E_{\rm inj}$,
  respectively. Middle panel: The bolometric SN light curves of the
  models from the left panel, compared to the data of SN 2017eaw from \citet{vandyk:19}.
  The black solid line shows the light curve obtained from the bare RSG profile with
  no energy injected at the base of its hydrogen envelope. Right panel: The photospheric velocity
  of the best fitting model ($E_{\rm inj}=5\times 10^{46}\,{\rm erg}$, $t_{\rm inj}=297.1\,{\rm d}$),
  as well as the velocity at the locations $\tau_{\rm Sob}=1.0$, $3.0$ and $5.0$ in this model.
  Black markers show the velocity measured from the Fe II $\lambda$5169 lines 
  of SN 2017eaw.} 
  \label{fig:fig3}
\end{figure*}

In Figure~\ref{fig:fig2}, we show the density profiles of the same
model as a function of radial coordinate, at different moments
of time after the energy injection. Before day $\sim$100, a weak shock wave 
can be seen propagating through the envelope. After it reaches the surface,
the outermost layers of the model start to expand. The black curve shows
the original RSG profile before the energy injection. 
We suggest that the wave heating may serve as a
theoretically justified mechanism for the formation of what we called
CSM in our earlier works \citep{morozova:17,morozova:18}.

\subsection{Effect on SN lightcurve}

With envelope density profiles as a function of time, we then model the
SN light curves as a function of $E_{\rm inj}$ and $t_{\rm inj}$.
We compare the obtained SN light curves to the bolometric 
light curve of SN 2017eaw and look for the 
best fitting parameters by minimizing the $\chi^2$.

The left panel of Figure~\ref{fig:fig3} shows $\chi^2/\chi^2_{\rm min}$ 
as a function of $E_{\rm inj}$ and $t_{\rm inj}$ in our model.
The gray shaded region on the left hand side
of the panel corresponds to the times when the weak shock wave launched by
the energy injection has not yet reached the stellar surface. 
During this time, no observable outburst is expected from the progenitor, 
and the density profile/SN light curve are not strongly altered.
The minimum $\chi^2$ (marked by the magenta circle) 
corresponds to $5\times 10^{46}\,{\rm erg}$ of energy being injected $\approx$297
days before the core-collapse. In this model, a visible outburst would
take place $\approx$197 days before the SN. For comparison, the blue and lime
circles mark some of the models in which the outburst happens at
$\approx$80 days and $\approx$160 before the SN, respectively.

\begin{figure*}
  \centering
  \includegraphics[width=0.95\textwidth]{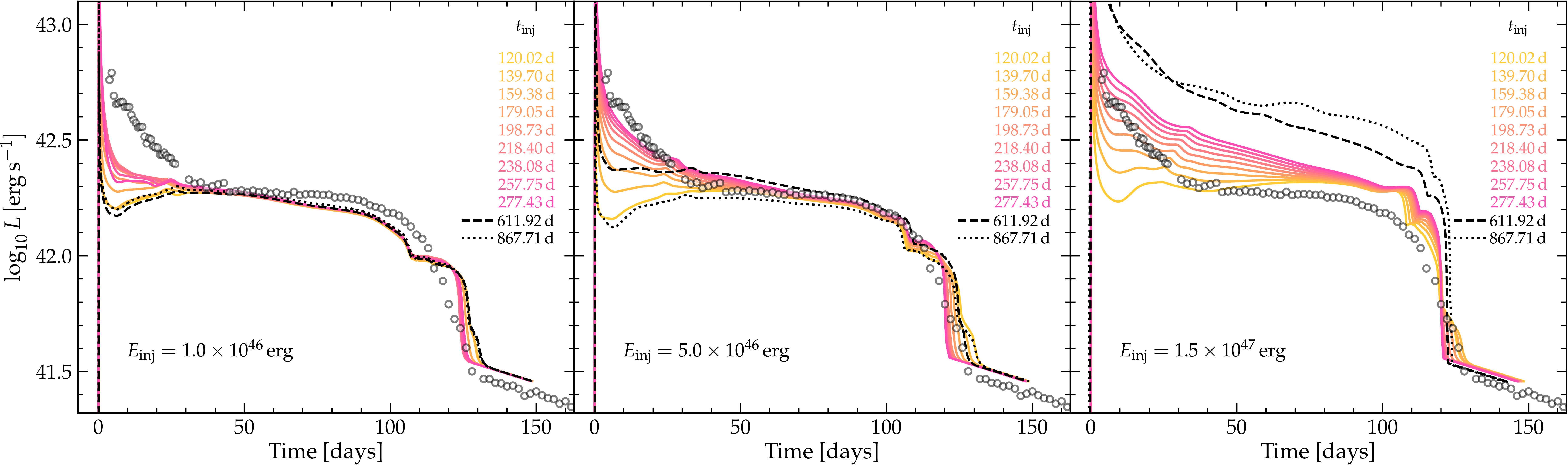}
  \caption{SN light curves of the KEPLER $M_{\rm ZAMS}=15\,M_{\odot}$ model, 
  obtained for the different values
  of $E_{\rm inj}$ and $t_{\rm inj}$. Gray markers show the data of SN 2017eaw.
  In the models with low $E_{\rm inj}$ (left and middle panels), 
  the density profile keeps expanding
  until day $\sim 300$ after the energy injection, and the SN light curves obtained
  from these models show progressive rise in the early bolometric luminosity. After
  day $\sim 300$, the ejected material decelerates and starts falling back onto the
  progenitor, reversing the progression of the SN light curves towards their initial state. 
  To avoid clutter in the plots, we show the continuous 
  evolution of the SN light curves until $t_{\rm inj}\approx 300\,{\rm d}$, while plotting the light curves
  for $t_{\rm inj}\approx 600\,{\rm d}$ and $t_{\rm inj}\approx 850\,{\rm d}$ in a different style. 
  For the high values of $E_{\rm inj}$, this reversal in the SN light curve progression 
  happens at much later times.} 
  \label{fig:fig4}
\end{figure*}

The bolometric light curves of these models are shown in the middle
panel of Figure~\ref{fig:fig3} with corresponding colors. The
black solid line plots the light curve obtained from the original
KEPLER $15\,M_{\odot}$ model without the energy injection, 
using the same set of SN parameters. Compared
to the black curve, all models with injected energy show the excess of
luminosity, especially in the first $\sim$40 days of the light curves. 
The magenta light curve demonstrates the best agreement with the 
data of SN 2017eaw.

For completeness, in the right panel of Figure~\ref{fig:fig3} we show the
velocity measured from the Fe II $\lambda$5169 lines of SN 2017eaw.
The solid magenta line gives the photospheric velocity of the 
minimum $\chi^2$ model, and one can see that this velocity significantly
underestimates the observed one. However, \citet{paxton:18}
resolved this issue by finding the velocity at the location where
the Sobolev optical depth, $\tau_{\rm Sob}$, of the Fe II $\lambda$5169 line 
is $\sim$1. Indeed, the value of $\tau_{\rm Sob}$ at the photosphere location
is much larger than 1 and reaches up to a few hundreds in
our models. The three dashed lines in the right panel of Figure~\ref{fig:fig3}
show the velocities at the locations where $\tau_{\rm Sob}=1.0$, $3.0$ and $5.0$
\footnote{To compute those, we use a table of the fraction of iron atoms in the
lower level of transition relevant for the Fe II $\lambda$5169 line, which is 
available in the public version of MESA and credited to Dan Kasen \citep{paxton:18}.}.
These velocities are in better agreement with the observations and able to
reproduce most of the measurements, except the ones at days $\sim$20 and $\sim$140.
At the same time, we note that pre-heating itself does not influence the
velocities significantly. The difference between the velocities of the
minimum $\chi^2$ model and the ones of the bare RSG model
is very small, and we don't show the latter in Figure~\ref{fig:fig3} to avoid
cluttering the plot.

Figure~\ref{fig:fig4} represents an overview of the SN light curves
obtained for the different values of $E_{\rm inj}$ and time $t_{\rm inj}$. It
is remarkable that even as little as $1.0\times 10^{46}\,{\rm erg}$ of energy
deposited into the envelope about a year before the core-collapse may have a 
significant impact on the final SN light curve. Therefore, if we want to reproduce the 
SN~IIP light curves with stellar evolution models, we cannot ignore
the impact of the late-stage nuclear burning on the progenitor structure. 
Ideally, the energy deposition should be modeled 
self-consistently during stellar evolution 
calculations (as was done, for example, by \citealt{fuller:17}), which we hope to
address in future works.

\begin{figure}
  \centering
  \includegraphics[width=0.41\textwidth]{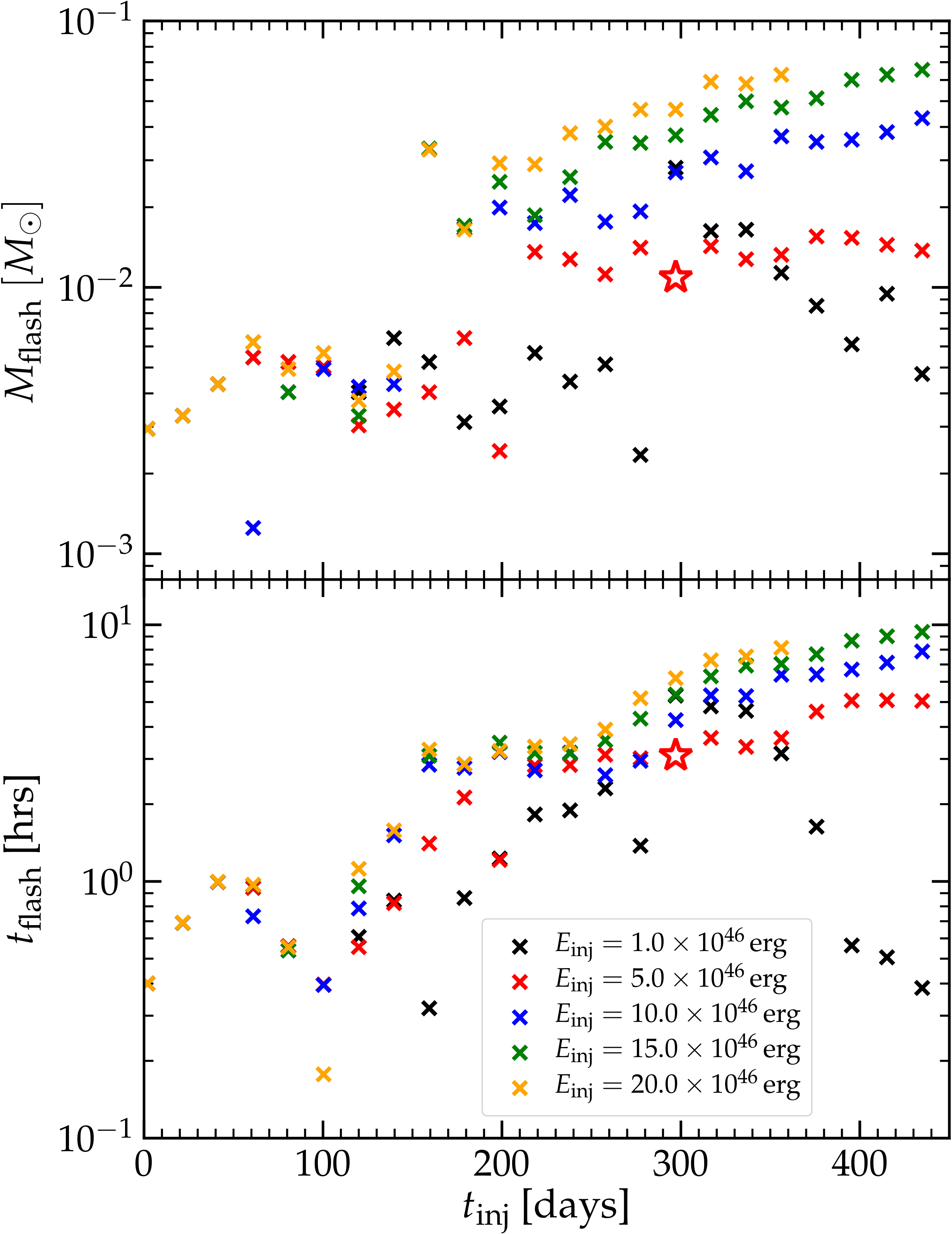}
  \caption{Top panel: The amount of mass that is expected to be flash-ionized
  by the shock breakout in our models, as a function of $t_{\rm inj}$ 
  for different energies $E_{\rm inj}$. Bottom panel: The time it takes for
  the flash-ionized material to be overtaken by the SN shock wave, after which the
  narrow emission lines in the spectra should disappear. The star marker in both
  panels corresponds to the best fitting model from Figure~\ref{fig:fig3}.} 
  \label{fig:fig5}
\end{figure}

Finally, we would like to emphasize the potential of our models to
explain one more aspect of the early emission from SNe~IIP, namely,
so called `flash-ionization' lines. These narrow emission lines appear in
the very early spectra of some SNe, and they are generally interpreted
as a sign of the shock wave interacting with the CSM \citep{khazov:16}.
These lines are generally short lived and disappear from the spectra
within just a few tens of hours \citep{yaron:17}.
SN 2017eaw also showed a weak narrow H$_{\alpha}$ emission line
in the spectrum taken at $\sim$1.4 days after the explosion, which did not
appear in the subsequent spectrum taken at $\sim$3.4 days post-explosion
\citep{rui:19}. In a recent study, \citet{dessart:17} has shown that a cocoon 
of material of only $\sim$0.01$\,M_{\odot}$ distributed out to about $5-10$ 
stellar radii is sufficient to reproduce the narrow lines seen in the early 
spectra of SNe 2013cu \citep{galyam:14} and 2013fs \citep{yaron:17}.
Recently, \citet{kochanek:18} suggested an alternative solution, explaining 
the flash spectroscopy by a collision interface formed between the 
regular stellar winds in a binary system.

In SNEC, we can pinpoint the moment of shock breakout from our
models based on the condition $\tau\approx c/v$, where $\tau$ is the
optical depth of the material above the shock, $v$ is the shock velocity,
and $c$ is the speed of light. The top panel of Figure~\ref{fig:fig5}
shows the mass of the material that lays above the shock at the moment
of breakout, denoted as $M_{\rm flash}$, 
as a function of $t_{\rm inj}$ for different $E_{\rm inj}$.
This material is expected to be flash-ionized by the UV radiation from the
shock breakout and emit narrow lines. The data plotted in Figure~\ref{fig:fig5}
appear noisy due to technical issues of determining the position and velocity 
of the shock wave, but the general trend can be seen very well. For small
$t_{\rm inj} < 100\,{\rm d}$, the shock breaks out at the very surface of the
models, illuminating only few $10^{-3}\,M_{\odot}$ of the downstream material.
Instead, at $t_{\rm inj} > 100\,{\rm d}$, at which the weak shock wave from the
energy injection could substantially pre-heat and 
expand the outer layers of the envelope,
the subsequent post-explosion shock wave breaks out deeper in mass
coordinate and illuminates more material, up to few $10^{-2}\,M_{\odot}$.
The bottom panel of Figure~\ref{fig:fig5} shows the time that it 
takes for the SN shock to overtake this material, $t_{\rm flash}$, 
after which the flash-ionization lines should disappear. We estimate this
time as $(R_{*}-R_{\rm br})/v_{\rm sh}$, where $R_{*}$ is the radius of the
model, $R_{\rm br}$ is the radial coordinate of the shock breakout, and 
$v_{\rm sh}$ is the shock velocity. 
In our models, $t_{\rm flash}$ constitutes several hours,
which is a factor of few short compared to observations. However, we anticipate that
more accurate simulations may be able to explain the flash-ionization lines
from SNe~IIP based on the outburst model suggested here.


\section{Conclusions and discussion}
\label{conclusions}

Our work demonstrates that even a moderate amount of energy
($10^{46}-10^{47}\,{\rm erg}$) deposited into 
RSG envelopes during late-stage nuclear 
burning can dramatically impact the final SN~IIP light curves. The
envelope expansion and mass ejection 
caused by this energy deposition resembles the dense CSM
that was used in previous works to explain the early luminosity excess
in the observed SN~IIP light curves compared to the bare RSG light curves
\citep{morozova:17,morozova:18}. 
The fact that such energy 
deposition is expected in RSGs due to wave energy transport
from the core to the envelope \citep{fuller:17} may explain the fact that
the vast majority of SNe~IIP benefit from the inclusion of some CSM in
their light curve models
\citep{morozova:18,moriya:18,das:18,paxton:18,foerster:18}. In addition,
the mass ejection caused by this energy deposition may explain the
formation of the narrow, flash-ionization lines observed in the
early spectra of many SNe~II \citep{khazov:16}.
Therefore, it is important to follow late-stage
nuclear burning in stellar evolution codes and ensure 
the proper energy exchange between
the core and the envelope by means of the wave transport.

A similar study to ours has been recently performed by \citet{ouchi:19},
where the authors investigated the influence of the late stage energy deposition 
into the RSG envelopes on their SN light curves. They 
found that a super-Eddington energy deposition changes the structure of
the RSG envelope so much that the resulting `light curves
and the evolution of photospheric velocity are all inconsistent with the 
observations of SNe~II'. However, we believe that the reason for this conclusion lies
in the long duration of the energy injection, which in their setup lasted 
for three years before the SN explosion. In this way, their models with 
high energy injection rates deposited too much energy into the envelopes,
while the models with low energy injection rates deposited the energy
too slowly to launch a weak shock wave and eject the outer layers of the
envelope. Instead, our setup was informed by the previous study of the
late-stage nuclear burning in RSGs \citep{fuller:17}, and we are 
able to demonstrate a good agreement between
our model and SN 2017eaw, which can be generalized to a larger
SN set in the future. At the same time, we note that any 
progenitor model whose density 
profile is similar to our best-fit model would match the data similarly 
well. Wave-driven outbursts may not be required if wind acceleration 
\citep{moriya:18} or some other envelope heating mechanism 
can support substantial CSM above the photosphere of stellar evolution 
models. 


Although our study does not provide quantitative predictions
of the pre-SN outburst light curves,
we can place constraints on their duration and amplitude.
Our results suggest that the observable outburst should take place not 
earlier than a few years before the SN. This roughly agrees with
the predictions of the wave heating model, in which wave heating
during core Ne burning can produce an outburst months to years
before core-collapse \citep{fuller:17}.
According to Figures 5 and 8 of \citet{fuller:17} (cases $\eta=1/3$ and $\eta=1$), 
$2\times 10^{46}\,{\rm erg}$ of energy deposited at the base of the RSG
envelope would lead to $\sim$20$\%$ increase in the progenitor luminosity
at the peak of the outburst, while the deposition of $6\times 10^{46}\,{\rm erg}$
would lead to the peak luminosity increase of a factor of 2. The duration
of the outburst is expected to be of the order of few months (see Figure 9
of \citealt{fuller:17}), after which both luminosity and effective temperature would
approach their initial values. Our work also demonstrates that it can take 
considerable time ($\sim$100 days or more) for the low energy shock 
driven by wave heating to reach the stellar surface. This means that 
when observations of pre-explosion outbursts (or at least constraints 
on their occurrence) are made, this timescale for traversing the star 
should be taken into account when making comparisons to specific 
stages of nuclear burning.

The optical Large Binocular Telescope (LBT) archival images of 
SN 2017eaw were analyzed by \citet{johnson:17}, 
where no signs of the progenitor pre-explosion activity were found. 
Further analysis of the Hubble Space Telescope (HST) data by 
\citet{rui:19} indicated a dimming of the
progenitor by $30\%$ one year before the explosion 
\citep[see also][]{vandyk:19}. It is
possible that the outburst could happen between the relatively scarce
HST measurements. The temperature evolution in
our simple models suggests that the ejected
material may quickly become optically thin, while not forming too much dust to
completely obscure the progenitor. In this view, it is important to calculate
better multi-band predictions of the outburst light curves,
which would provide a clearer picture of the outburst signatures that should be
present in pre-explosion data. This will be especially helpful once LSST \citep{abell:09}
is online, providing a long baseline of pre-explosion history for SN progenitors.

From \citet{vandyk:19}, the progenitor of SN 2017eaw may have
exhibited a moderate 
($\sim 40\%$) increase in the infrared Spitzer flux $1.6\,{\rm yr}$ before the explosion
(although the measurement uncertainty is still too large to claim it to be a definite
outburst). Another possibility is that an outburst occurred in the last $\sim$40 
days before the SN explosion, where there is no Spitzer data. 
However, Keck and Palomar images taken days before the explosion do not
show strong evidence for an outburst \citep{tinyanont:19}, so we disfavor this
possibility. Our current best model with $E_{\rm inj}=5\times 10^{46}\,{\rm erg}$
(expected $\sim 50-60\%$ increase in the luminosity at the outburst peak)
corresponds to the outburst happening $\sim$0.5$\,{\rm yr}$ before the
SN, which could not be captured by the available ground- and space-based observations. 
However, we emphasize that we have not carried out an extensive parameter
survey, and it may be possible to find other stellar models
(e.g., of different mass or radius) with different energy injection parameters
that can fit both the pre-SN variability constraints and the SN light curve.

\acknowledgments
V.M. acknowledges helpful discussions with David Radice, James Stone
and Adam Burrows.
A.L.P. acknowledges financial support for this research from a Scialog 
award made by the Research Corporation for Science Advancement.
This research is funded in part by a Rose Hills Innovator Grant, and by
grant HST-AR-15021.001-A. Numerical simulations were performed 
using Della cluster of Princeton University, as well as computing
services of the Institute for Computational and Data Sciences of the
Pennsylvania State University.

\bibliographystyle{apj}

\begin{thebibliography}{32}
\expandafter\ifx\csname natexlab\endcsname\relax\def\natexlab#1{#1}\fi

\bibitem[{{Buta} \& {Keel}(2019)}]{buta:19}
{Buta}, R.~J., \& {Keel}, W.~C. 2019, \mnras, 487, 832

\bibitem[{{Das} \& {Ray}(2017)}]{das:18}
{Das}, S., \& {Ray}, A. 2017, \apj, 851, 138

\bibitem[{{Dessart} {et~al.}(2017){Dessart}, {John Hillier}, \&
  {Audit}}]{dessart:17}
{Dessart}, L., {John Hillier}, D., \& {Audit}, E. 2017, \aap, 605, A83

\bibitem[{{F{\"o}rster} {et~al.}(2018){F{\"o}rster}, {Moriya}, {Maureira},
  {Anderson}, {Blinnikov}, {Bufano}, {Cabrera-Vives}, {Clocchiatti}, {de
  Jaeger}, {Est{\'e}vez}, {Galbany}, {Gonz{\'a}lez-Gait{\'a}n}, {Gr{\"a}fener},
  {Hamuy}, {Hsiao}, {Huentelemu}, {Huijse}, {Kuncarayakti}, {Mart{\'{\i}}nez},
  {Medina}, {Olivares E.}, {Pignata}, {Razza}, {Reyes}, {San Mart{\'{\i}}n},
  {Smith}, {Vera}, {Vivas}, {de Ugarte Postigo}, {Yoon}, {Ashall}, {Fraser},
  {Gal-Yam}, {Kankare}, {Le Guillou}, {Mazzali}, {Walton}, \&
  {Young}}]{foerster:18}
{F{\"o}rster}, F., {Moriya}, T.~J., {Maureira}, J.~C., {et~al.} 2018, Nature
  Astronomy

\bibitem[{{Fuller}(2017)}]{fuller:17}
{Fuller}, J. 2017, \mnras, 470, 1642

\bibitem[{{Gal-Yam} {et~al.}(2014){Gal-Yam}, {Arcavi}, {Ofek}, {Ben-Ami},
  {Cenko}, {Kasliwal}, {Cao}, {Yaron}, {Tal}, {Silverman}, {Horesh}, {De Cia},
  {Taddia}, {Sollerman}, {Perley}, {Vreeswijk}, {Kulkarni}, {Nugent},
  {Filippenko}, \& {Wheeler}}]{galyam:14}
{Gal-Yam}, A., {Arcavi}, I., {Ofek}, E.~O., {et~al.} 2014, \nat, 509, 471

\bibitem[{{Georgy} {et~al.}(2013){Georgy}, {Ekstr{\"o}m}, {Granada}, {Meynet},
  {Mowlavi}, {Eggenberger}, \& {Maeder}}]{georgy:13}
{Georgy}, C., {Ekstr{\"o}m}, S., {Granada}, A., {et~al.} 2013, \aap, 553, A24

\bibitem[{{Johnson} {et~al.}(2018){Johnson}, {Kochanek}, \&
  {Adams}}]{johnson:17}
{Johnson}, S.~A., {Kochanek}, C.~S., \& {Adams}, S.~M. 2018, \mnras, 480, 1696

\bibitem[{{Khazov} {et~al.}(2016){Khazov}, {Yaron}, {Gal-Yam}, {Manulis},
  {Rubin}, {Kulkarni}, {Arcavi}, {Kasliwal}, {Ofek}, {Cao}, {Perley},
  {Sollerman}, {Horesh}, {Sullivan}, {Filippenko}, {Nugent}, {Howell}, {Cenko},
  {Silverman}, {Ebeling}, {Taddia}, {Johansson}, {Laher}, {Surace},
  {Rebbapragada}, {Wozniak}, \& {Matheson}}]{khazov:16}
{Khazov}, D., {Yaron}, O., {Gal-Yam}, A., {et~al.} 2016, \apj, 818, 3

\bibitem[{{Kochanek}(2019)}]{kochanek:18}
{Kochanek}, C.~S. 2019, \mnras, 483, 3762

\bibitem[{{Kochanek} {et~al.}(2017){Kochanek}, {Fraser}, {Adams}, {Sukhbold},
  {Prieto}, {M{\"u}ller}, {Bock}, {Brown}, {Dong}, {Holoien}, {Khan},
  {Shappee}, \& {Stanek}}]{kochanek:17}
{Kochanek}, C.~S., {Fraser}, M., {Adams}, S.~M., {et~al.} 2017, \mnras, 467,
  3347

\bibitem[{{LSST Science Collaboration} {et~al.}(2009){LSST Science
  Collaboration}, {Abell}, {Allison}, {Anderson}, {Andrew}, {Angel}, {Armus},
  {Arnett}, {Asztalos}, {Axelrod}, {Bailey}, {Ballantyne}, {Bankert},
  {Barkhouse}, {Barr}, {Barrientos}, {Barth}, {Bartlett}, {Becker}, {Becla},
  {Beers}, {Bernstein}, {Biswas}, {Blanton}, {Bloom}, {Bochanski}, {Boeshaar},
  {Borne}, {Bradac}, {Brandt}, {Bridge}, {Brown}, {Brunner}, {Bullock},
  {Burgasser}, {Burge}, {Burke}, {Cargile}, {Chand rasekharan}, {Chartas},
  {Chesley}, {Chu}, {Cinabro}, {Claire}, {Claver}, {Clowe}, {Connolly}, {Cook},
  {Cooke}, {Cooray}, {Covey}, {Culliton}, {de Jong}, {de Vries}, {Debattista},
  {Delgado}, {Dell'Antonio}, {Dhital}, {Di Stefano}, {Dickinson}, {Dilday},
  {Djorgovski}, {Dobler}, {Donalek}, {Dubois-Felsmann}, {Durech},
  {Eliasdottir}, {Eracleous}, {Eyer}, {Falco}, {Fan}, {Fassnacht}, {Ferguson},
  {Fernandez}, {Fields}, {Finkbeiner}, {Figueroa}, {Fox}, {Francke}, {Frank},
  {Frieman}, {Fromenteau}, {Furqan}, {Galaz}, {Gal-Yam}, {Garnavich},
  {Gawiser}, {Geary}, {Gee}, {Gibson}, {Gilmore}, {Grace}, {Green}, {Gressler},
  {Grillmair}, {Habib}, {Haggerty}, {Hamuy}, {Harris}, {Hawley}, {Heavens},
  {Hebb}, {Henry}, {Hileman}, {Hilton}, {Hoadley}, {Holberg}, {Holman},
  {Howell}, {Infante}, {Ivezic}, {Jacoby}, {Jain}, {R}, {Jedicke}, {Jee},
  {Garrett Jernigan}, {Jha}, {Johnston}, {Jones}, {Juric}, {Kaasalainen},
  {Styliani}, {Kafka}, {Kahn}, {Kaib}, {Kalirai}, {Kantor}, {Kasliwal},
  {Keeton}, {Kessler}, {Knezevic}, {Kowalski}, {Krabbendam}, {Krughoff},
  {Kulkarni}, {Kuhlman}, {Lacy}, {Lepine}, {Liang}, {Lien}, {Lira}, {Long},
  {Lorenz}, {Lotz}, {Lupton}, {Lutz}, {Macri}, {Mahabal}, {Mandelbaum},
  {Marshall}, {May}, {McGehee}, {Meadows}, {Meert}, {Milani}, {Miller},
  {Miller}, {Mills}, {Minniti}, {Monet}, {Mukadam}, {Nakar}, {Neill}, {Newman},
  {Nikolaev}, {Nordby}, {O'Connor}, {Oguri}, {Oliver}, {Olivier}, {Olsen},
  {Olsen}, {Olszewski}, {Oluseyi}, {Padilla}, {Parker}, {Pepper}, {Peterson},
  {Petry}, {Pinto}, {Pizagno}, {Popescu}, {Prsa}, {Radcka}, {Raddick},
  {Rasmussen}, {Rau}, {Rho}, {Rhoads}, {Richards}, {Ridgway}, {Robertson},
  {Roskar}, {Saha}, {Sarajedini}, {Scannapieco}, {Schalk}, {Schindler},
  {Schmidt}, {Schmidt}, {Schneider}, {Schumacher}, {Scranton}, {Sebag},
  {Seppala}, {Shemmer}, {Simon}, {Sivertz}, {Smith}, {Allyn Smith}, {Smith},
  {Spitz}, {Stanford}, {Stassun}, {Strader}, {Strauss}, {Stubbs}, {Sweeney},
  {Szalay}, {Szkody}, {Takada}, {Thorman}, {Trilling}, {Trimble}, {Tyson}, {Van
  Berg}, {Vand en Berk}, {VanderPlas}, {Verde}, {Vrsnak}, {Walkowicz}, {Wand
  elt}, {Wang}, {Wang}, {Warner}, {Wechsler}, {West}, {Wiecha}, {Williams},
  {Willman}, {Wittman}, {Wolff}, {Wood-Vasey}, {Wozniak}, {Young}, {Zentner},
  \& {Zhan}}]{abell:09}
{LSST Science Collaboration}, {Abell}, P.~A., {Allison}, J., {et~al.} 2009,
  arXiv e-prints, arXiv:0912.0201

\bibitem[{{Moriya} {et~al.}(2018){Moriya}, {F{\"o}rster}, {Yoon},
  {Gr{\"a}fener}, \& {Blinnikov}}]{moriya:18}
{Moriya}, T.~J., {F{\"o}rster}, F., {Yoon}, S.-C., {Gr{\"a}fener}, G., \&
  {Blinnikov}, S.~I. 2018, \mnras, 476, 2840

\bibitem[{{Moriya} {et~al.}(2017){Moriya}, {Yoon}, {Gr{\"a}fener}, \&
  {Blinnikov}}]{moriya:17}
{Moriya}, T.~J., {Yoon}, S.-C., {Gr{\"a}fener}, G., \& {Blinnikov}, S.~I. 2017,
  \mnras, 469, L108

\bibitem[{{Morozova} {et~al.}(2015){Morozova}, {Piro}, {Renzo}, {Ott},
  {Clausen}, {Couch}, {Ellis}, \& {Roberts}}]{morozova:15}
{Morozova}, V., {Piro}, A.~L., {Renzo}, M., {et~al.} 2015, \apj, 814, 63

\bibitem[{{Morozova} {et~al.}(2017){Morozova}, {Piro}, \&
  {Valenti}}]{morozova:17}
{Morozova}, V., {Piro}, A.~L., \& {Valenti}, S. 2017, \apj, 838, 28

\bibitem[{{Morozova} {et~al.}(2018){Morozova}, {Piro}, \&
  {Valenti}}]{morozova:18}
---. 2018, \apj, 858, 15

\bibitem[{{Ofek} {et~al.}(2013){Ofek}, {Sullivan}, {Cenko}, {Kasliwal},
  {Gal-Yam}, {Kulkarni}, {Arcavi}, {Bildsten}, {Bloom}, {Horesh}, {Howell},
  {Filippenko}, {Laher}, {Murray}, {Nakar}, {Nugent}, {Silverman}, {Shaviv},
  {Surace}, \& {Yaron}}]{ofek:13}
{Ofek}, E.~O., {Sullivan}, M., {Cenko}, S.~B., {et~al.} 2013, \nat, 494, 65

\bibitem[{{Ofek} {et~al.}(2014){Ofek}, {Sullivan}, {Shaviv}, {Steinbok},
  {Arcavi}, {Gal-Yam}, {Tal}, {Kulkarni}, {Nugent}, {Ben-Ami}, {Kasliwal},
  {Cenko}, {Laher}, {Surace}, {Bloom}, {Filippenko}, {Silverman}, \&
  {Yaron}}]{ofek:14}
{Ofek}, E.~O., {Sullivan}, M., {Shaviv}, N.~J., {et~al.} 2014, \apj, 789, 104

\bibitem[{{O'Neill} {et~al.}(2019){O'Neill}, {Kotak}, {Fraser}, {Sim},
  {Benetti}, {Smartt}, {Mattila}, {Ashall}, {Callis}, {Elias-Rosa},
  {Gromadzki}, \& {Prentice}}]{oneill:18}
{O'Neill}, D., {Kotak}, R., {Fraser}, M., {et~al.} 2019, \aap, 622, L1

\bibitem[{{Ouchi} \& {Maeda}(2019)}]{ouchi:19}
{Ouchi}, R., \& {Maeda}, K. 2019, \apj, 877, 92

\bibitem[{{Paxton} {et~al.}(2018){Paxton}, {Schwab}, {Bauer}, {Bildsten},
  {Blinnikov}, {Duffell}, {Farmer}, {Goldberg}, {Marchant}, {Sorokina},
  {Thoul}, {Townsend}, \& {Timmes}}]{paxton:18}
{Paxton}, B., {Schwab}, J., {Bauer}, E.~B., {et~al.} 2018, \apjs, 234, 34

\bibitem[{{Quataert} {et~al.}(2016){Quataert}, {Fern{\'a}ndez}, {Kasen},
  {Klion}, \& {Paxton}}]{quataert:16}
{Quataert}, E., {Fern{\'a}ndez}, R., {Kasen}, D., {Klion}, H., \& {Paxton}, B.
  2016, \mnras, 458, 1214

\bibitem[{{Quataert} \& {Shiode}(2012)}]{quataert:12}
{Quataert}, E., \& {Shiode}, J. 2012, \mnras, 423, L92

\bibitem[{{Rui} {et~al.}(2019){Rui}, {Wang}, {Mo}, {Xiang}, {Zhang}, {Maund},
  {Gal-Yam}, {Wang}, \& {Zhang}}]{rui:19}
{Rui}, L., {Wang}, X., {Mo}, J., {et~al.} 2019, \mnras, 485, 1990

\bibitem[{{Shiode} \& {Quataert}(2014)}]{shiode:14}
{Shiode}, J.~H., \& {Quataert}, E. 2014, \apj, 780, 96

\bibitem[{{Smartt}(2009)}]{smartt:09}
{Smartt}, S.~J. 2009, \araa, 47, 63

\bibitem[{{Sukhbold} {et~al.}(2016){Sukhbold}, {Ertl}, {Woosley}, {Brown}, \&
  {Janka}}]{sukhbold:16}
{Sukhbold}, T., {Ertl}, T., {Woosley}, S.~E., {Brown}, J.~M., \& {Janka}, H.-T.
  2016, \apj, 821, 38

\bibitem[{{Tinyanont} {et~al.}(2019){Tinyanont}, {Kasliwal}, {Krafton}, {Lau},
  {Rho}, {Leonard}, {De}, {Jencson}, {Mawet}, {Millar-Blanchaer}, {Nilsson},
  {Yan}, {Gehrz}, {Helou}, {Van Dyk}, {Serabyn}, {Fox}, \&
  {Clayton}}]{tinyanont:19}
{Tinyanont}, S., {Kasliwal}, M.~M., {Krafton}, K., {et~al.} 2019, \apj, 873,
  127

\bibitem[{{Van Dyk}(2017)}]{vandyk:17}
{Van Dyk}, S.~D. 2017, {Supernova Progenitors Observed with HST}, ed. A.~W.
  {Alsabti} \& P.~{Murdin}, 693

\bibitem[{{Van Dyk} {et~al.}(2019){Van Dyk}, {Zheng}, {Maund}, {Brink},
  {Srinivasan}, {Andrews}, {Smith}, {Leonard}, {Morozova}, {Filippenko},
  {Conner}, {Milisavljevic}, {de Jaeger}, {Long}, {Isaacson}, {Crossfield},
  {Kosiarek}, {Howard}, {Fox}, {Kelly}, {Piro}, {Littlefair}, {Dhillon},
  {Wilson}, {Butterley}, {Yunus}, {Channa}, {Jeffers}, {Falcon}, {Ross},
  {Hestenes}, {Stegman}, {Zhang}, \& {Kumar}}]{vandyk:19}
{Van Dyk}, S.~D., {Zheng}, W., {Maund}, J.~R., {et~al.} 2019, \apj, 875, 136

\bibitem[{{Yaron} {et~al.}(2017){Yaron}, {Perley}, {Gal-Yam}, {Groh}, {Horesh},
  {Ofek}, {Kulkarni}, {Sollerman}, {Fransson}, {Rubin}, {Szabo}, {Sapir},
  {Taddia}, {Cenko}, {Valenti}, {Arcavi}, {Howell}, {Kasliwal}, {Vreeswijk},
  {Khazov}, {Fox}, {Cao}, {Gnat}, {Kelly}, {Nugent}, {Filippenko}, {Laher},
  {Wozniak}, {Lee}, {Rebbapragada}, {Maguire}, {Sullivan}, \&
  {Soumagnac}}]{yaron:17}
{Yaron}, O., {Perley}, D.~A., {Gal-Yam}, A., {et~al.} 2017, Nature Physics, 13,
  510

\end{thebibliography}

\end{document}